\documentclass[%
 reprint,
 amsmath,amssymb,
 aps,
]{revtex4-2}

\usepackage{graphicx}
\usepackage{dcolumn}
\usepackage{bm}

\usepackage{graphicx}
\usepackage{amsmath}
\usepackage{xr}
\usepackage{pifont}
\usepackage{amssymb}
\usepackage{float}
\usepackage{xcolor}
\usepackage{mathtools}
\usepackage{tikz}
\usepackage{physics}
\usepackage{mathtools}
\usepackage{epsfig}
\usepackage{calculus}
\usepackage{calc}
\usetikzlibrary{calc,angles,quotes}
\usepackage{pgfplots}
\pgfplotsset{compat=1.12} 
\usepgfplotslibrary{ternary}

\pgfplotsset{/pgfplots/colormap={colmap}{
        rgb255=(0,0,102)
        rgb255=(0,0,255)
        rgb255=(0,255,255)
        rgb255=(255,255,255)
        rgb255=(255,255,0)
        rgb255=(255,0,0)
        rgb255=(102,0,0)
    }
}

\begin{document}

\preprint{APS/123-QED}

\title{Entanglement Propagation in Integrable Heisenberg Chains from a New Lens}

\author{Peyman Azodi}
\email{pazodi@princeton.edu}
\author{Herschel A.Rabitz}%

\affiliation{%
 Department of Chemistry, Princeton University, Princeton, New Jersey 08544, USA 
}%




\date{\today}

\begin{abstract}
 The exact single-magnon entanglement evolution in Heisenberg chains is obtained using the Quantum Correlation Transfer Function (QCTF) formulation. A dual, i.e., frequency and time-domain, analysis shows that the transient dynamics of individual spins' entanglement is described via a Bessel function of the first kind. Through QCTF, we bypass the evaluation of the full system's state for the purpose of obtaining entanglement. Although it is known that the observable entanglement edge is formed by the arrival of a stream of quasi-particles that travel with the maximum group velocity, we show how the \textit{early} quasi-particles travel \textit{faster than the maximum group velocity} of the chain and contribute to entanglement production. Our results can be extended to the multi-magnon regime, therefore opening up the means to better interpret equilibration dynamics and thermodynamics in Heisenberg chains.

\end{abstract}

\maketitle


\par\textit{Introduction.-} {Understanding entanglement propagation in non-equilibrium many-body quantum systems is valuable for both fundamental and practical reasons, especially given new developments regarding the interrelation between entanglement and thermodynamics \cite{ langen2013local,doi:10.1126/science.aaf6725,lewis2019dynamics,gogolin2016equilibration,rigol2008thermalization,srednicki1994chaos,deutsch1991quantum,PhysRevLett.126.160602}. { From the viewpoint of quantum information sciences, it is crucial to understand how entanglement propagates, as it determines the rates at which correlations can be transported in quantum circuits \cite{epstein2017quantum}.} Integrable quantum systems have been a prime subject in the study of entanglement dynamics due to their importance and algebraic structure \cite{ calabrese2016introduction,doi:10.1126/science.288.5465.475, PhysRevLett.96.136801,calabrese2005evolution, calabrese2020entanglement, alba2017entanglement,bertini2022growth,PhysRevB.101.094304}.  A method to determine the long-term behavior of entanglement and general local observables is by using the Eigenstate Thermalization Hypothesis (ETH), which conjunctures that these quantities can be derived from conventional thermodynamic ensembles at appropriate effective temperatures. \cite{deutsch1991quantum,  goldstein2006canonical,PhysRevE.50.888,eisert2015quantum,Essler_2016}. For instance, Generalized Gibbs Ensembles (GGE) have been used to predict the \textit{asymptotic} behavior of observables \cite{rigol2007relaxation, barthel2008dephasing, calabrese2011quantum, cazalilla2012thermalization,  kormos2014analytic}, including both successful \cite{collura2013equilibration, ilievski2015complete, wouters2014quenching, piroli2016exact} and unsuccessful \cite{pozsgay2014correlations, mestyan2015quenching} cases of this formulation in studying the Heisenberg model. Although there have been analytical advancements in understanding entanglement, determining the precise time evolution and the fundamental mechanism of the equilibration process generally requires numerical methods, with only a few exceptions \cite{PasqualeCalabrese_2004,PhysRevLett.123.210601,bertini2022entanglement}. }
\par In relativistic quantum systems with short-length interactions, Lieb and Robinson's theorem provides a bound based on the maximum group velocity and a resulting causality light-cone for the ballistic propagation of correlations, beyond which they must decay exponentially \cite{lieb1972finite,bravyi2006lieb}. This phenomenon has been experimentally observed in several instances \cite{cheneau2012light, jurcevic2014quasiparticle}. The presence of long-range interactions breaks the Lieb-Robinson bound, but further modifications can be made to obtain the correlation transport velocity \cite{hauke2013spread,schachenmayer2013entanglement, cevolani2018universal,schneider2021spreading, foss2015nearly}, which has proven to remain finite under certain circumstances \cite{chen2019finite}.
Moreover, the spread of correlations is shown to have a double causality structure with different velocities, where one case corresponds to the edge - which is faster, given by the phase velocity in the lattice - and the other case is associated with the extremum of correlation transport \cite{cevolani2018universal, despres2019twofold}. 
\par Although entanglement propagation in integrable Heisenberg chains is viewed to be a mature subject, this paper brings additional insight gained \textit{through a new lens}, the Quantum Correlation Transfer Functions (QCTFs) \cite{QCTF,azodi2022directly}. In this framework, the dynamical properties of a subsystem's entanglement are encoded in the residues of a complex (QCTF) function which can be calculated directly from the system's Hamiltonian and its pre-quench state. To this end, the dynamics of entanglement is quantified using a geometric measure: the squared area spanned by projected wave functions (onto a local basis for the subsystem of interest) \cite{QCTF}. In the case of subsystems with two-energy levels, this measure of entanglement reduces to the determinant of the reduced density matrices in the Laplace domain. In order to obtain the QCTF function, we assign a unique integer number to an arbitrary set of eigenstates for the underlying Hilbert space; nevertheless, the residues of the QCTF, which encode entanglement between subsystems, are invariant to the chosen basis. 
\par This treatment enables a full analysis of the single-magnon entanglement quench \textit{dynamics} in ferromagnetic Heisenberg spin-$\frac{1}{2}$ chains of arbitrary length. The choice of local quench based on a single-magnon excitation allows for the study of the velocity of propagation of correlations in the chain. In addition to the exact characterization of entanglement dynamics in the chain with an arbitrary number of spins, another main finding of this paper is the exact calculation of the early entanglement edge velocity $v_e=\frac{e}{2}v_{group}$ in anisotropic Heisenberg chains, with no dependence on the anisotropy in the chain. This finding shows that early entanglement is transferred through the chain with a velocity faster than the group velocity. However, the substantial entanglement build-up happens upon the accumulation of quasi-particles, which leads to an observable entanglement edge, numerically found to travel with the group velocity \cite{cevolani2018universal}. Our results add to the understanding of entanglement dynamics in this well-studied class of integrable systems by revealing new aspects of this phenomenon through the use of QCTF. As will be discussed in the Summary and Conclusion Section, the QCTF formulation offers a unique perspective on entanglement propagation which complements the commonly used methods.

\par \textit{Model and QCTF Analysis.-} The goal of this paper is to study the quench entanglement dynamics of a single-magnon state in an anisotropic Heisenberg chain with the following Hamiltonian,
\begin{equation}
    \mathbf{H}=-\mathbf{J}\sum _{j=-\frac{N-1}{2}}^{\frac{N-1}{2}}\big({S}^x_j {S}^x_{j+1}+{S}^y_j {S}^y_{j+1}+\Delta ({S}^z_j {S}^z_{j+1}-\frac{1}{4})\big ),
\end{equation}
with $N$ (odd) being the number of spins and with periodic boundary condition, i.e., ${S}_{\frac{N+1}{2}}={S}_{-\frac{N-1}{2}}$.  Here, $\mathbf{J}$ and $\Delta$ denote the interaction strength and anisotropy. The pre-quench state of the quantum chain is the single-magnon excitation of one of the degenerate ferromagnetic states (${S}_j^-\ket{F}=0$), where ${S}^\pm_j=S^x_j\pm iS^y_j$ are the spin raising/lowering operators at site $j$. Without loss of generality due to transnational invariance, we choose the magnon state as $\ket{0}\doteq\mathbf{S}_0^+ \ket{F}
$. Since the product state $\ket{F}$ is an eigenstate of the Hamiltonian, the local quench in the magnon excitation is exclusively responsible for the entanglement evolution in the chain. In what follows, we study how entanglement evolves and propagates through the chain, using the QCTF formulation.
\par The QCTF treatment starts by labeling an arbitrary set of basis kets for the one-magnon sector (with $\expval{S^Z_{total}}= \expval{\sum {S}^z}=- \frac{N}{2}+1$) as $\ket{p}\doteq S_p^+\ket{F}$, where $p=-\frac{N-1}{2},\cdots, 0, \cdots, \frac{N-1}{2}$. Since $[\mathbf{H}, S^Z_{total}]=0$, the higher-order magnon sectors can be ignored in the resolvent function, defined as $\mathbf{G}(s)=(s\mathbf{I}-\frac{i}{\hbar}\mathbf{H})^{-1}$ ( $s$ is the Laplace variable). Employing the translational invariance of the chain, the coordinate Bethe ansatz gives the eigenstates of the chain as $\ket{K}=(N)^{-\frac{1}{2}}\sum_{p=-\frac{N-1}{2}}^{\frac{N-1}{2}} e^{ipK}\ket{p}$, with the dispersion relation $E(K)=\mathbf{J}(\Delta-\cos(K))$ and momenta $K=\frac{2\pi}{N}m$; $m=-\frac{N-1}{2}, \cdots, \frac{N-1}{2}$.Therefore, the resolvent defined in the sub-Hilbert space of interest (i.e., one-magnon sector) can be written as follows:
\begin{equation}\label{res}
\begin{split}
    \mathbf{G}(s)=\frac{1}{N}\sum_{m=-\frac{N-1}{2}}^{\frac{N-1}{2}}\Bigg[& \Bigg(s-\frac{i\mathbf{J}}{\hbar}\bigg(\Delta-\cos(\frac{2\pi}{N}m)\bigg)\Bigg )^{-1}\\&\sum_{p_1,p_2}e^{\frac{2im\pi}{N}(p_1-p_2)}\ket{p_1}\bra{p_2}\Bigg].
\end{split}\end{equation}

\par
\begin{figure}
    \centering
\includegraphics[width = 0.55\textwidth]{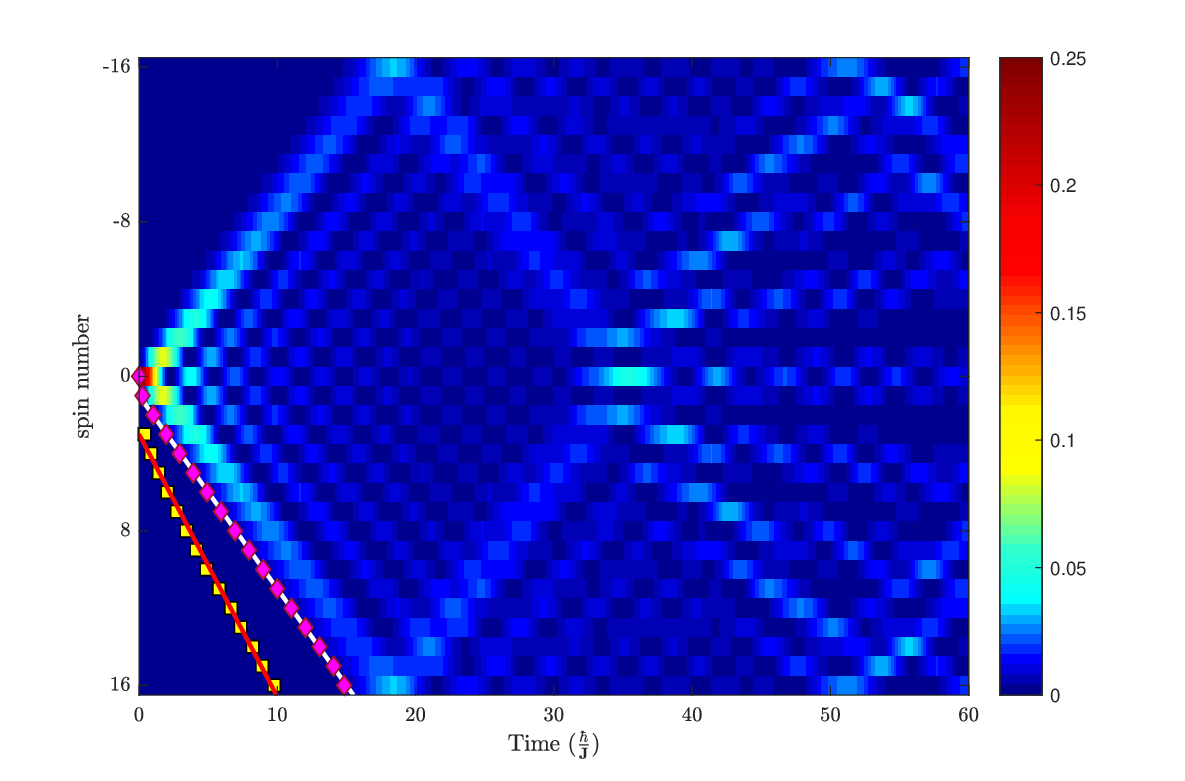}
    \caption{Exact evolution of entanglement after a local quench at $t=0$ in the middle of the chain obtained using the QCTF formulation. Due to periodic boundary conditions and translational symmetry, all of the spins can be considered to be in the middle of the chain. The evolution shows clear light-cone behavior until quasi-particles arrive at the middle of the chain (located at the upper and lower edges of the figure). {The yellow squares (violet diamonds) represent when the entanglement measure for each spin surpasses the value of $\nu=1.8e-6$ ($
\nu=1.5e-3$). The path of the yellow squares indicates the early entanglement edge, which moves at the velocity $\frac{e}{2}\mathbf{J}$, predicted theoretically and illustrated by the red solid line. The violet diamonds illustrate the observable entanglement edge, advancing with the group velocity, depicted by the solid white line.}}
    \label{fig:my_label0}
\end{figure}

 In the QCTF framework, entanglement dynamics of each individual spin can be obtained by finding the residues of a corresponding QCTF transformation. The first step is to find the QCTF centered at the spin number $q$  (the subsystem of interest). For this model, the QCTF is defined as \cite{QCTF}:
\begin{equation}\label{Q1}
    \mathcal{K}_q(s,z_d,z_a)=\sum_{\substack{p=-\frac{N-1}{2}\\p\neq q}}^{\frac{N-1}{2}} z_d^{q-p}z_a^{q+p}\mel{0}{\mathbf{G}^\dagger(s^*)}{p}\star \mel{q}{\mathbf{G}(s)}{0},
\end{equation}
where $z_a,z_d ${$\in \mathbb{C}$} are variables and the operator $\star$ is the convolution in the $s$ domain and regular multiplication in $z_d$ and $z_a$ domains \cite{QCTF}. In the remainder of the paper, a basic application of this operator, namely $(s+i\omega_1)^{-1}\star(s+i\omega_2)^{-1}=(s+i(\omega_1+\omega_2))^{-1}$ will be used. Inserting (\ref{res}) in the QCTF (\ref{Q1}) leads to
\begin{widetext}
\begin{equation}\label{400}
    \mathcal{K}_q=\frac{1}{N^2}\sum_{\substack{m_1,m_2,p=-\frac{N-1}{2}\\p\neq q}}^{\frac{N-1}{2}}{z_d^{q-p}z_a^{q+p}\Bigg(s-\frac{i\mathbf{J}}{\hbar}\bigg(\cos(\frac{2\pi}{N}m_2)-\cos(\frac{2\pi}{N}m_1)\bigg)\Bigg )^{-1} e^{\frac{2i\pi}{N}(m_1q-m_2p)}}.
\end{equation}
\end{widetext}
This formula can be understood as a three-variable transformation of the density matrix: two transformations with one parallel to the diagonal ($z_d$) and the other perpendicular to the diagonal ($z_a$) array of elements of the density matrix, as well as a transformation to the Laplace domain ($s$), which reflects the time-evolution of entanglement. Note that the dependence on $\Delta$ is not present in the QCTF function. This variable is a constant shift in the energy of each fixed-magnon block on the diagonal of the Hamiltonian, therefore it does not affect the linear combination of eigenvalues that appears in the QCTF entanglement measure. {The independence of entanglement dynamics upon $\Delta$ is only applicable when $\Delta$ is finite.} Having determined the QCTF, the dynamical entanglement measure of spin $q$ ($\Tilde{\mathcal{Q}}_q(s)$) can be obtained using the following relation \cite{QCTF}:

\begin{equation}\label{main}
\begin{split}
    \Tilde{\mathcal{Q}}_q(s)=&\underset{{\substack{z_d=0\\z_a=0}}}{\mathbf{Res}}\big((z_dz_a)^{-1}{\mathcal{K}_q(z_d,z_a,s)\star \mathcal{K}_q^*(1/z^*_d,1/z^*_a,s^*)} \big)\\&-{\mathcal{K}_{d}(s)\star\mathcal{K}_{d}^*(s^*)},
    \end{split}
\end{equation}
with $\mathcal{K}_d(s)=\eval{\underset{{\substack{z_d=0}}}{\mathbf{Res}}\big(z_d^{-1}{\mathcal{K}_q(z_d,z_a,s)}\big )}_{z_a=1}$. One can show that $\Tilde{\mathcal{Q}}_q(s)$ is the determinant of the reduced density matrix of spin $q$, in the Laplace domain \cite{QCTF}. These residues can easily be found upon expanding the $\star$ multiplication in (\ref{main}) using (\ref{400}), which gives the following dynamical entanglement measure:
\begin{widetext}
\begin{equation}\label{ent}
    \Tilde{\mathcal{Q}}_q(s)=\frac{1}{N^4}\sum_{\substack{m_1,m_2\\m_3,m_4\\p\neq q}}
    \Bigg( s-\frac{i\mathbf{J}}{\hbar}\bigg(\cos(\frac{2\pi}{N}m_2)-\cos(\frac{2\pi}{N}m_1)-\cos(\frac{2\pi}{N}m_4)+\cos(\frac{2\pi}{N}m_3)\bigg)\Bigg )^{-1} e^{\frac{2i\pi}{N}\big(q(m_1-m_3)+p(m_4-m_2)\big)}.
\end{equation}
\end{widetext}
This equation provides the frequency spectrum of the dynamical entanglement of spin $q$. By taking the inverse Laplace transform, one finds the entanglement time-evolution of each spin, which is shown in Figure \ref{fig:my_label0} for $N=33$ spins. Note that the poles of this function (\ref{ent}) have the inversion symmetries $m_i \leftrightarrow -m_i$ and also the $m_1 \leftrightarrow m_4$, and $m_2 \leftrightarrow m_3$ symmetries.

\begin{figure*}
    \centering

    \includegraphics[width = 0.48\textwidth]{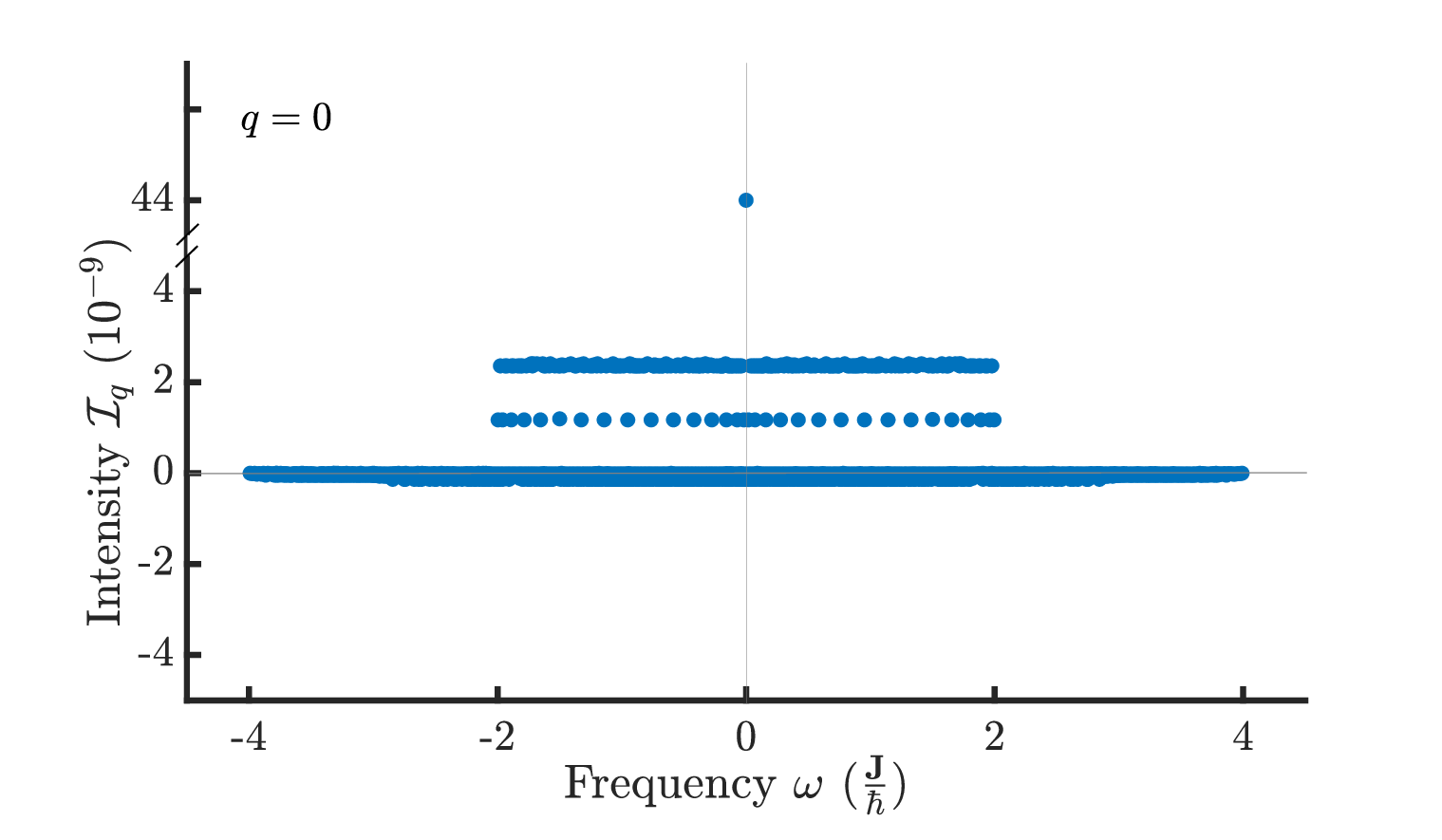}
    \includegraphics[width = 0.48\textwidth]{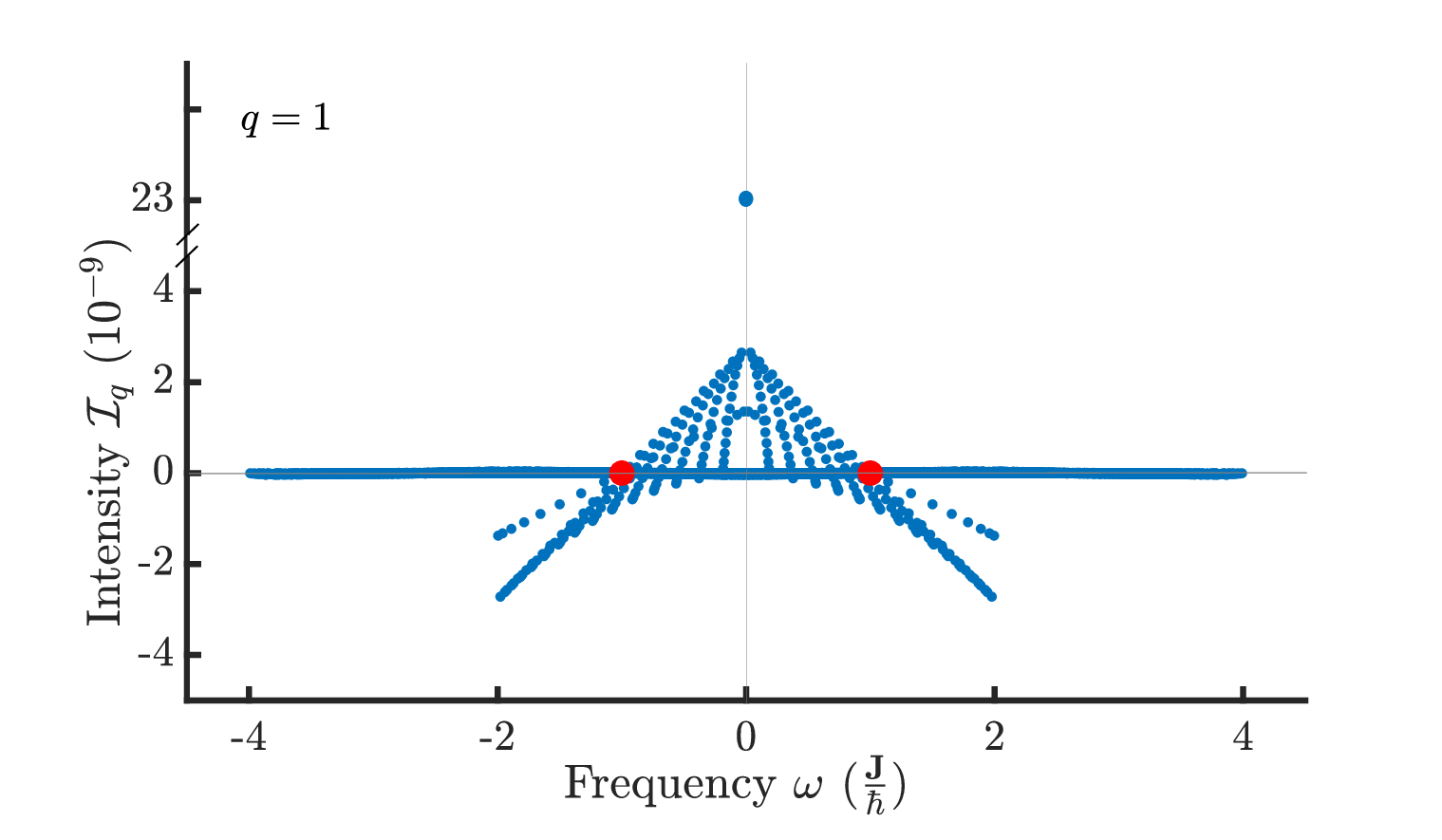}
   \includegraphics[width = 0.48\textwidth]{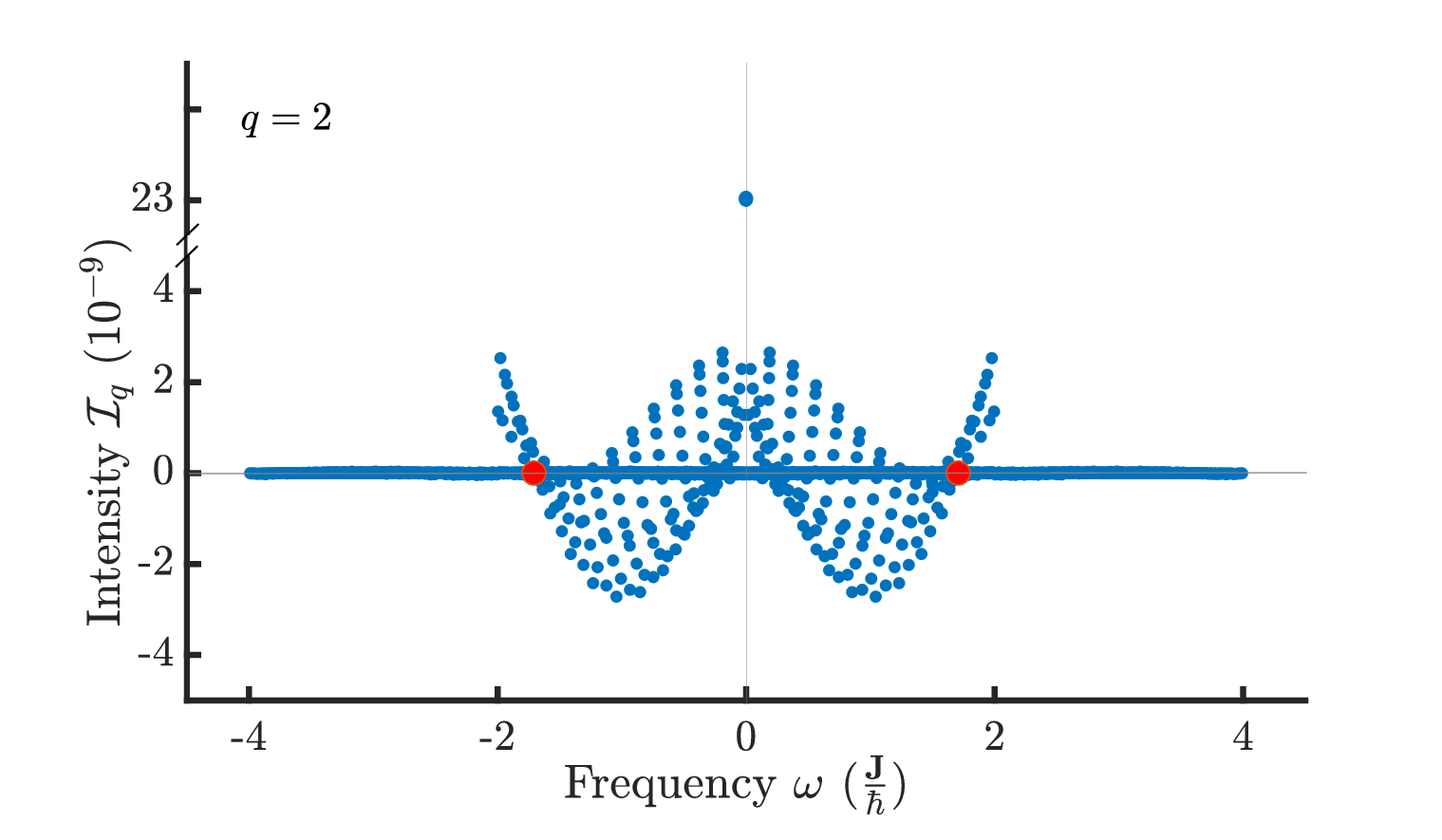}
   \includegraphics[width = 0.48\textwidth]{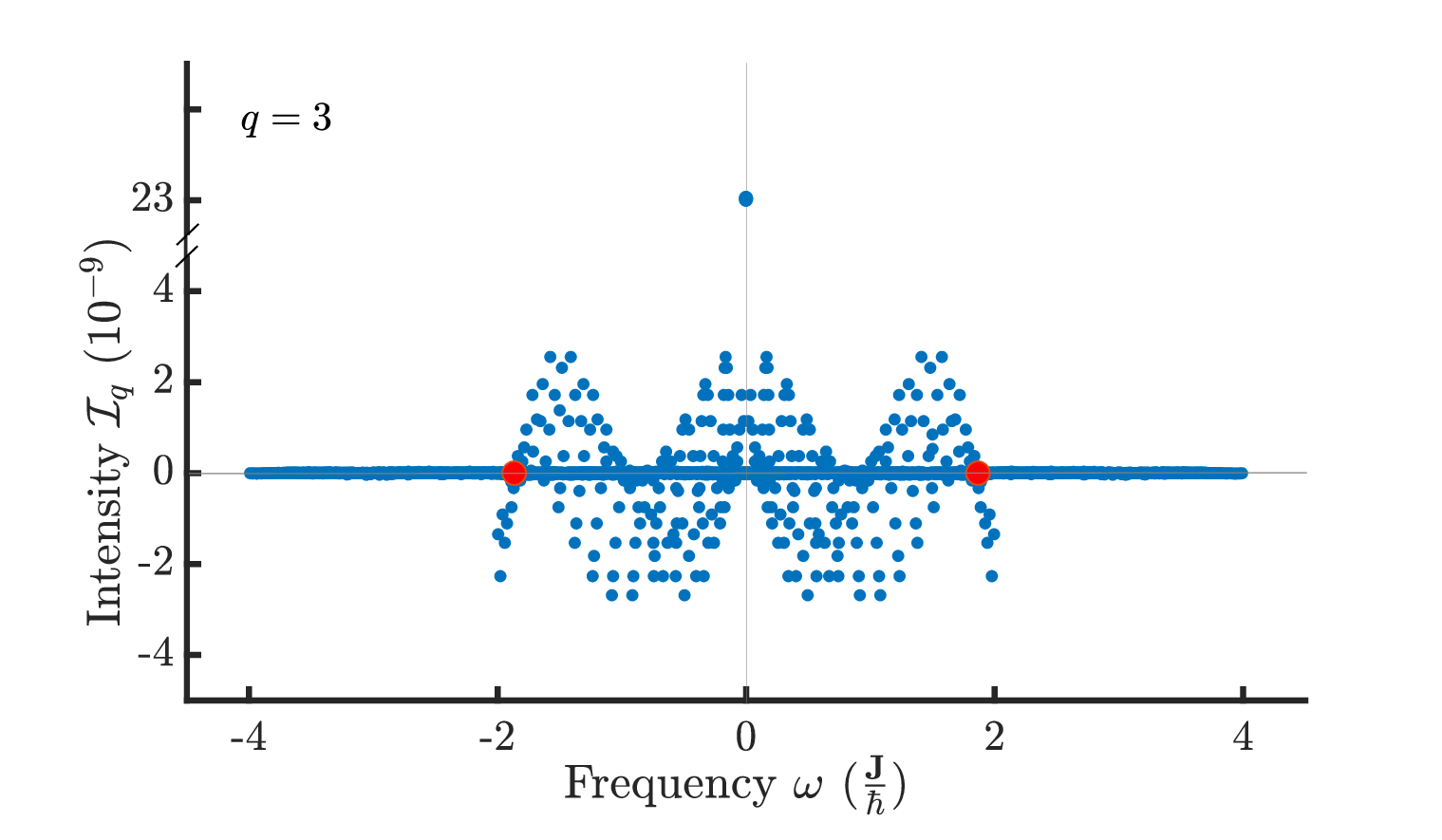}
   \includegraphics[width = 0.48\textwidth]{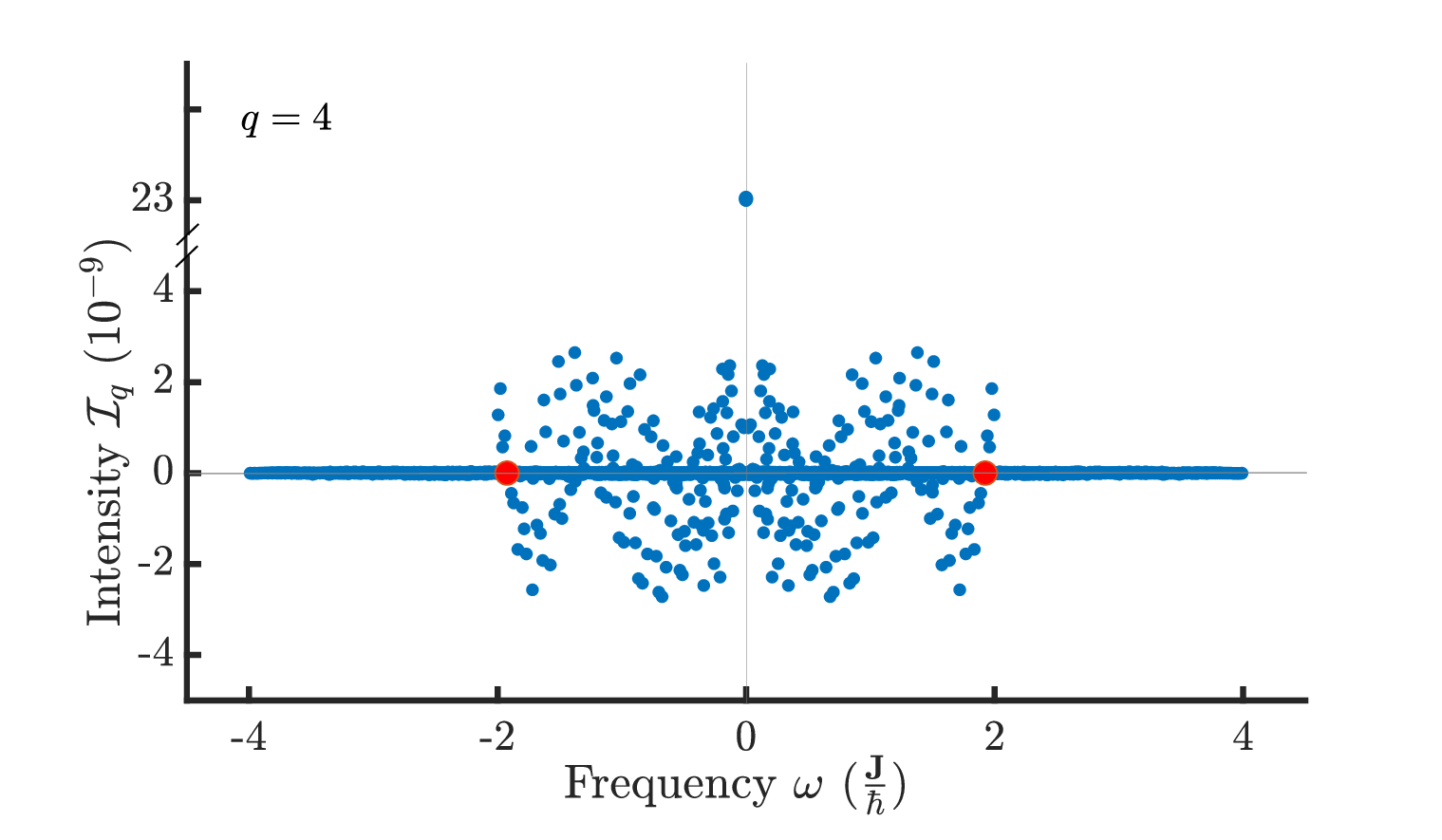}
   \includegraphics[width = 0.48\textwidth]{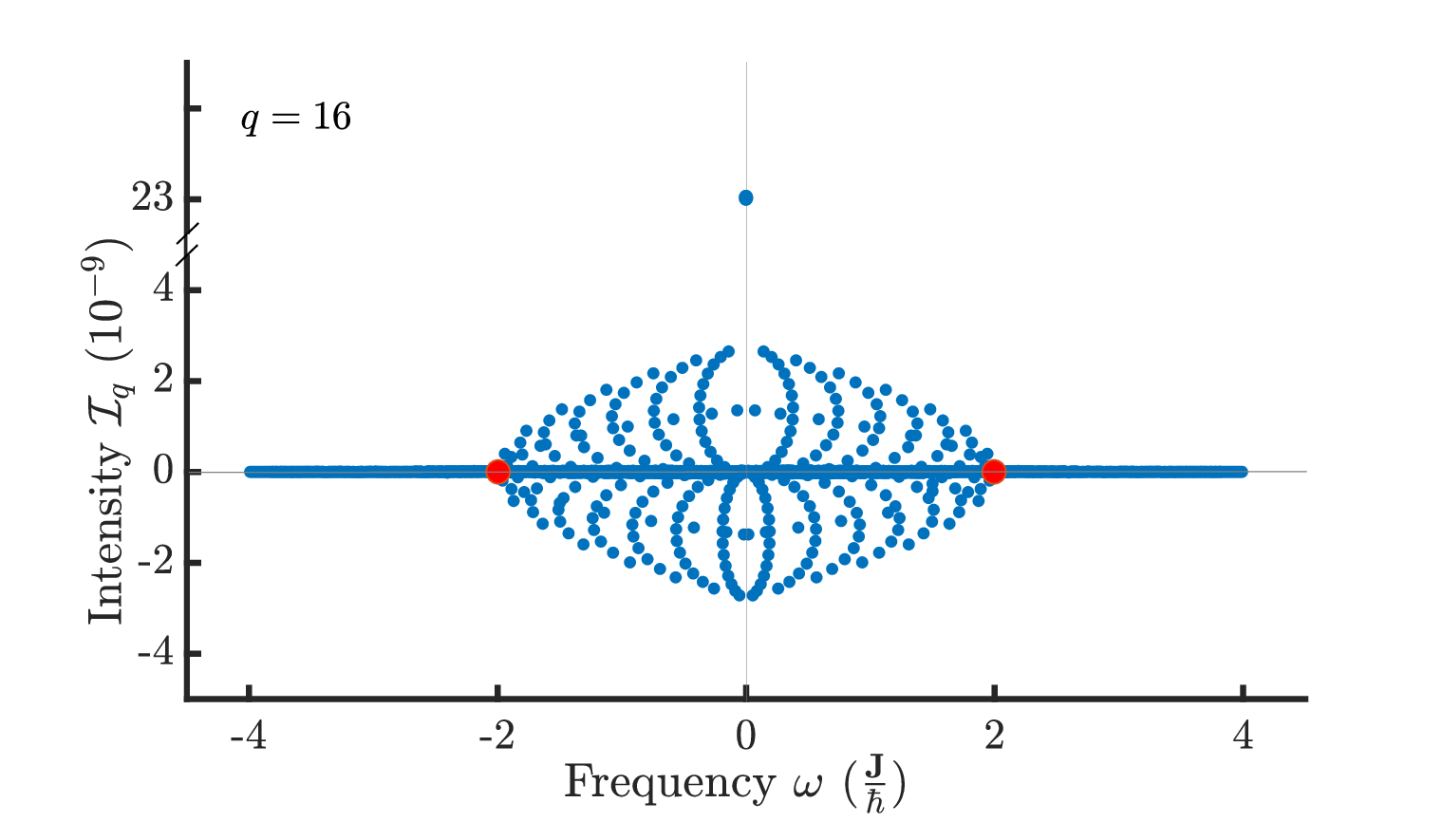}
  
    \caption{{Entanglement frequency spectra for different spins in the chains, starting from $q=0$ (the initially excited spin) on the upper left, and other spins $q$ ($q$ denotes the distance of the spin to the initially quenched spin). The spectra are given in (\ref{intt}-\ref{110}), and entail two main classes of poles: dominant poles, that contribute mainly to the entanglement evolution, and suppressed poles with intensities close to $\mathcal{I}_q=0$ (parallel to the horizontal axis). For the initially excited spin ($q=0$), the dominant frequency spectrum consists of two equi-intensity lines of poles (refer to the main text for explanation), as illustrated by horizontal lines of poles. For further spins (higher $q$'s), the frequency spectra feature a string of poles close to the cut-off frequency ($\omega=\frac{2 \mathbf{J}}{\hbar}$) that approach zero in intensity at $\approx 2\mathbf{J} \cos ^2(\frac{pi}{4q})$ (these poles are marked by red circles). }  }
    \label{Fig_spectra}
\end{figure*}

\par \textit{Analysis-} We will present two different analyses of equation (\ref{ent}), respectively in the frequency and time domains. The relation (\ref{ent}) shows that the entanglement frequency components (poles of $\Tilde{\mathcal{Q}}_q(s)$) must be upper-bounded by $|s|<\frac{4\mathbf{J}}{\hbar}$. Importantly, as will be demonstrated, the frequencies in the upper half region, $\frac{2\mathbf{J}}{\hbar}<|s|<\frac{4\mathbf{J}}{\hbar}$, are highly suppressed, polynomially in $N$, and therefore negligible in the $N \rightarrow \infty$ limit. To produce these frequencies on the higher end of the spectrum, cooperative addition of all four terms in the frequency argument (i.e., $\cos(\frac{2\pi}{N}m_i)$) is required, which necessarily demands that $m_2\neq m_4$. In this case, the inner summation over $p$ will lead to:
\begin{equation}
    \sum _{p\neq q}e^{\frac{2i\pi}{N}\big(q(m_1-m_3)+p(m_4-m_2)\big)}=-e^{\frac{2i\pi}{N}q(m_1-m_3+m_4-m_2)}.
\end{equation}
Therefore, the inner summation reduces to a number with unit norm. Given the $N^{-4}$ scaling in $\tilde{\mathcal{Q}}_q(s)$, this situation not only suppresses all of the higher end frequencies ($\frac{\mathbf{2J}}{\hbar}<|s|<\frac{4\mathbf{J}}{\hbar}$), but also the majority of frequencies at the lower end ($0<|s|<\frac{2\mathbf{J}}{\hbar}$). As a result, the dominant frequencies correspond to $m_2=m_4$.
\par Entanglement of the initially excited spin (i.e., $q=0$) can be obtained directly from (\ref{ent}). In this case, the intensity of the dominant frequencies is proportional to their abundance. Thus, finding the intensity of each frequency component in the entanglement measure entails counting the instances when each particular frequency emerges as the four-tuple $(m_1, m_2, m_3, m_4)$ varies. This statement follows since the exponential term becomes unity when $q=0$ and $m_2-m_4=0$. As a result, the entanglement frequency spectrum of the initially excited spin ($q=0$) consists of two equal-intensity lines (see Figure \ref{Fig_spectra}), one for $m_1 m_3\neq 0$, and one for $m_1 m_3=0$, with lower intensity due to a lower number count.
\par Analogously, for the general case of $q\neq 0$, the dynamical measure (\ref{ent}) gives the propagation of entanglement throughout the chain. Here, the transient behavior of entanglement (corresponding to the fast time scales) is of main interest. Transient features of entanglement correspond to the poles close to $|s|\approx \frac{2\mathbf{J}}{\hbar}$, which can be verified to correspond to the following (note that all frequencies appear in positive and negative pairs. Here only positive frequencies are considered for brevity):
\begin{equation}
    m_2=m_4, (m_1,m_3)\approx (0,\pm \pi).
\end{equation}
Therefore, the fastest dominant peak corresponds to:
\begin{equation}
    (m_1,m_3)=(0, \pm \frac{N-1}{2}).
\end{equation}
By employing the new set of variables $\epsilon=|m_1|-|m_3|-\frac{N}{2}$ and $\delta=|m_1|+|m_3|-\frac{N}{2}$, the intensity ($\mathcal{I}_q$) and frequency ({$\omega$}) of the (non-zero) dominant peaks are:
\begin{eqnarray}\label{intt}
        &\mathcal{I}_q(\epsilon,\delta) \propto (-1)^q\Big (\cos (\frac{2\pi q}{N}\epsilon)+\cos (\frac{2\pi q}{N}\delta)\Big ) \label{100},
        \\ &{\omega}(\epsilon,\delta)=\frac{2\mathbf{J}}{\hbar}\cos(\frac{\pi}{N}\epsilon)\cos(\frac{\pi}{N}\delta)\label{110}.
\end{eqnarray}
As a result, based on (\ref{100}), one expects to observe a \textit{string} of poles, close to and below the cut-off frequency ($\frac{2\mathbf{J}}{\hbar}$), the intensity of which decay to zero (and cross the horizontal axis in Figure \ref{Fig_spectra}) more rapidly as $q$ increases. A simple calculation shows that these crossings of the zero intensity line occur each time $|m_1|$ or $|m_3|$ crosses the pole near $\frac{N}{2}-\frac{N}{4q}$ and $\frac{N}{4q}$. Therefore, the first (meaning closest to the cut-off frequency) crossing corresponds to $m_1=\pm \frac{N-1}{2}$ and $|m_3|\approx \frac{N}{4q}$, which, according to (\ref{110}), will be at the frequency (shown with red marks in Figure \ref{Fig_spectra}):
\begin{equation}
   2\mathbf{J}\cos(\frac{\pi}{4q}-\frac{\pi}{2N}) \cos(\frac{\pi}{4q}+\frac{\pi}{2N})\approx 2\mathbf{J}\cos ^2 (\frac{\pi}{4q}).
\end{equation}
This mechanism filters out fast modes through the oscillatory behavior of poles near the cut-off frequency, leading to retarded growth of entanglement for farther spins (larger $q$). This behavior is analyzed in detail in the following paragraphs. 
\par Here, we present an alternative and more in-depth analysis of the transient entanglement dynamics in the time domain. We demonstrate that entanglement of spins at the $q$'th distance from the initially quenched spin obeys the transient behavior $\sim (\frac{v_et}{q})^{2q}$, where $v_e$ is a universal constant of the chain and describes the velocity of propagation for the entanglement edge. Given the even symmetry of the frequency components, the inverse Laplace transform of (\ref{ent}), which gives the entanglement dynamics of spin $q$, has the following general form in the time domain (since only $t\geq 0$ is our focus, we consider the symmetrized version of entanglement measure (i.e., around $t=0$ with $\frac{\mathcal{Q}_q(t)+\mathcal{Q}_q(-t)}{2}$)):
\begin{equation}
    \mathcal{Q}_q(t)=\sum_{j}{\mathcal{I}_{q,j} \cos(\omega_j t)},
\end{equation}
where $j$ is the index for all possible frequencies ($\omega _j=-is_j$, with $s_j$ being the poles of (\ref{ent})), arising from the four-tuples $(m_1,m_2,m_3,m_4)$, and $\mathcal{I}_{q,j}$ is the intensity corresponding to $\omega_j$, when considering spin $q$. Therefore, all of the odd derivatives (with respect to time) of $\mathcal{Q}_q(t)$ at $t=0$ vanish, and the even derivatives are:
\begin{equation}
    \eval{\mathcal{Q}_q^{(2r)}(t)}_{t=0}=(-1)^{r} \sum_{j}{\mathcal{I}_{q,j} \omega_j^{2r}}.
\end{equation}
We define the vector  $\mathbf{\mathcal{I}}^q\doteq [\mathcal{I}_{q,1},\cdots, \mathcal{I}_{q,j},\cdots]^T$, thus, given that all of the first even (up to $2(q-1)$) derivatives of $\mathcal{Q}_q(t)$ vanish at $t=0$ we have the linear system of equations:
\begin{equation}\label{vanish}
    V \mathcal{I}^q=\mathbf{0}
\end{equation}
where $V$ is the following transposed Vandermonde matrix:
\begin{equation}\label{Vande}
    (V_{kj})\doteq (i\omega_j)^{2(k-1)}; k=1,\cdots,q.
\end{equation}


Thus, the intensities in the entanglement dynamics of the $q$'th spin, i.e., $\mathcal{I}^q$, should belong to the null space of $V$. The proof of this statement can be found in the supporting material. Moreover, it is shown that the higher (than $2(q-1)$) order derivatives of $\mathcal{Q}_q(t)$, denoted by $\mathcal{Q}^{(2(q+\Bar{k}))}_q(t)$, are:
\begin{equation}\label{deriv}\begin{split}
    \eval{Q_q^{(2(q+\Bar{k}))}(t)}_{t=0}=&(\frac{\mathbf{J}}{2\hbar})^{2(q+\Bar{k})} {(-1)^{\Bar{k}} }\binom{2(q+\Bar{k})}{\Bar{k},q,q+\Bar{k}} \\ &\times{\,_2F_1(-\Bar{k},-\Bar{k}-q;q+1;1)},\end{split}
\end{equation}
where $\,_{2}F_{1}$ denotes the Gaussian hypergeometric function. Note that this expression is exact for $0\leq \Bar{k} <q$ and, for larger $\Bar{k}$'s, the contribution from suppressed poles in the entanglement frequency spectrum starts to emerge. The contributions of the suppressed poles are polynomially small in the entanglement's transient behavior. 


To study the transient behavior of entanglement, the lower order terms in the Taylor series expansion of (\ref{deriv}) can be used. By transient, we refer to the evolution of entanglement before the quasi-particles reach the $q$th distant spin, wherein entanglement is exponentially small in $q$, per the Lieb-Robinson theorem. For this purpose, The entanglement measure $ Q_q(t)$ can be rewritten in its asymptotic Taylor expansion form, which reveals an important feature of the entanglement dynamics: faster than group velocity propagation of entanglement in the system. Based on the fact that all lower (than $2q$th) derivatives of $ Q_q(t)$ vanish, and using the Stirling's approximation, $q!\approx \sqrt{2\pi q} (q/e)^q$, the leading term ($\Bar{k}=0$) in the Taylor expansion of $ Q_q(t)$ that governs the transport behavior of entanglement is:
\begin{equation}\begin{split}\label{www}
    \frac{1}{(q!)^2}(\frac{\mathbf{J}t}{2\hbar})^{2q}\approx \frac{1}{2\pi q}(\frac {t}{\tau_q})^{2q}; \tau _q=\frac{2q\hbar}{\mathbf{J}e}.
\end{split}    
\end{equation}
Accordingly, the following is the speed at which the edge of early entanglement travels down the chain:
\begin{equation}
 v_e= \frac{e\mathbf{J}}{2\hbar}=\frac{e}{2}v_g. 
\end{equation}
Therefore, $\tau_q=\frac {2q\hbar}{e \mathbf{J}}$ is the time before which entanglement is exponentially small in $q$ for $q$th distant spin from the initially excited spin (see equation (\ref{www})). As we expect, $\tau_q$ depends linearly on $q$, when $q\gg 1$.
{ Here, we introduce a new  entanglement growth time-scale that coincides to the \textit{early entanglement edge}. This velocity, given by $\frac{e}{2}\mathbf{J}$, exceeds the group velocity. The early entanglement edge corresponds to when the entanglement at each site surpasses a value $\nu$, which is significantly smaller than the typical entanglement values but larger than the tunneling effects. The explicit form of the early entanglement edge growth is given by (\ref{www}). If $\nu$ is comparable to the typical entanglement values, then the \textit{observable entanglement edge} is captured. The observable entanglement edge transports with the group velocity, as will be explained next.} By considering all orders of the Taylor series, the transient entanglement evolution, given by the hypergeometric function (\ref{deriv}), can be approximated as follows (see the supplementary material for derivation):
\begin{equation}\label{bes}
    Q_q^{transient}(t)\approx \alpha_q J_{2q}(\frac{2\mathbf{J}t}{\hbar}),
\end{equation}
where $J_{2q}$ is a Bessel function of the first kind, of order $2q$ and $\alpha_q\doteq (4^{-q})\binom{2q}{q}$. We should note that since in (\ref{bes}), all terms in the Taylor series of $ Q_q(t)$ are used, in comparison with (\ref{www}) where only the leading term is used, this equation provides an enhanced approximation of the transient entanglement evolution up to time $\frac{\hbar
}{\mathbf{J}}$, which is beyond the entanglement edge ($\tau_q$). Based on (\ref{bes}), the main body of quasi-particles, that constitute the observable entanglement edge, travel with the group velocity, $\frac{\mathbf{J}}{\hbar}$. Nevertheless, equation (\ref{www}) shows that early quasi-particles travel faster, i.e., with velocity $\frac{e}{2\hbar}\mathbf{J}$.  
\par \textit{Summary and Conclusion -} In this paper, we presented new insights into the well-studied Heisenberg spin chain model. Through this new QCTF lens, a detailed and meticulous analysis was made possible that revealed new aspects of out-of-equilibrium entanglement transport in Heisenberg chains from the system's Hamiltonian, without directly calculating the system's time evolution. We fully obtained the evolution of entanglement in an arbitrarily long Heisenberg spin chain after a local quench. Moreover, the QCTF allowed for a detailed analysis in the frequency domain, in addition to an accompanying time-domain analysis, which revealed the velocity of the early entanglement edge in this class of spin chains. This velocity, in addition to providing fundamental insight into the mechanism of correlation transport in Heisenberg chains, is of practical importance to quantum information processing technologies. Most significantly, this velocity prescribes the fastest rate at which information can be transported beyond quantum tunneling effects in quantum networks with similar effective dynamics.
\par The QCTF formulation provided a fresh approach to entanglement propagation, which complements the existing methods used to study this important problem. Here, we emphasize what this new method offers, in the general context of the presented analysis. Firstly, although the studied system is integrable, QCTF, as a generic method, is not limited to considering entanglement in integrable systems. The common approach to treating entanglement propagation is through obtaining different velocities from the system's dispersion relation. This method assumes that the propagation of entanglement is monochromatic, through which the notions of group and phase velocity are well-defined. However, this feature is not guaranteed in all scenarios and initial conditions, as in the newly emerging subject of out-of-equilibrium quantum phenomena \cite{defenu2023outofequilibrium}. In the present work, the QCTF method allowed for a more direct study of entanglement, where the analysis was appropriately confined to only the relevant portion of the Hilbert space explored by the system's state. This gave rise to the new understanding of faster-than-maximum group velocity propagation of entanglement. \\Although the spectrum of the integrable system studied in this paper is given by Bethe' ansatz, there is a major simplification enabled through QCTF, which is the \textit{direct} determination of entanglement. With QCTF, the specific features of the system, giving rise to entanglement propagation, are targeted for analysis; Whereas in the well-known exact-diagonalization scheme, one would first obtain the full time evolution of the system's wave function, calculate the sub-system's reduced density matrix, and eventually find entanglement. Therefore, even though similar conclusions may potentially be obtained using exact-diagonalization methods, the QCTF approach provided a significantly more straightforward route. Moreover, the applicability of the QCTF formulation goes beyond systems for which exact diagonalization is feasible.  
\par A natural direction for future QCTF research entails consideration of multi-magnon entanglement evolution in Heisenberg XXZ chains, which should provide insights into the non-equilibrium quantum statistics in this class of system. Moreover, by implementing the ``string hypothesis" of the Beth\'e ansatz in the QCTF framework, entanglement dynamics in Heisenberg chains can be studied in a variety of global quench settings, such as the tilted-ferromagnetic state \cite{alba2017entanglement}. Ultimately, we hope to use the QCTF framework to study the problem of entanglement propagation in Heisenberg chains with long-range interactions {with focus on different measures of correlation and in the presence of a thermal bath \cite{dehghani2020entanglement}}. This class of systems is not integrable and therefore requires further effort to treat (e.g., employing time-independent perturbation theory). 

\begin{acknowledgments}
P.A acknowledges support from the Princeton Program in Plasma Science and Technology (PPST). H.R and P.A acknowledges support from the U.S Department Of Energy (DOE) grant (DE-FG02-02ER15344).
\end{acknowledgments}


\bibliography{citations}

\end{document}